\documentclass[aps,nofootinbib,twocolumn,prl,superscriptaddress,amsmath,amssymb,amsfonts,floatfix]{revtex4-2}
\usepackage{amsmath}
\usepackage{amssymb}
\usepackage{amsfonts}
\usepackage[dvips]{graphicx}
\usepackage{epsfig}
\usepackage{tabularx}
\usepackage{color}
\usepackage{siunitx}
\usepackage{orcidlink}

\begin{document}

\title{Cooperative stabilization of persistent currents in superfluid ring networks}

\author{M. Ciszak~\orcidlink{0000-0003-1087-6105}}
	\affiliation{CNR-INO Sede Secondaria di Sesto Fiorentino - c/o LENS, Via Nello Carrara, 1 - 50019 Sesto Fiorentino, Italy}

\author{N. Grani~\orcidlink{0000-0001-6107-9726}}
\affiliation{CNR-INO Sede Secondaria di Sesto Fiorentino - c/o LENS, Via Nello Carrara, 1 - 50019 Sesto Fiorentino, Italy}
\affiliation{European Laboratory for Nonlinear Spectroscopy (LENS), Via N. Carrara 1, 50019 Sesto Fiorentino, Italy}
\affiliation{INFN, Sezione di Firenze, 50019 Sesto Fiorentino, Italy}
\author{D. Hernández-Rajkov~\orcidlink{0009-0002-1908-4227}}
\affiliation{CNR-INO Sede Secondaria di Sesto Fiorentino - c/o LENS, Via Nello Carrara, 1 - 50019 Sesto Fiorentino, Italy}
\affiliation{European Laboratory for Nonlinear Spectroscopy (LENS), Via N. Carrara 1, 50019 Sesto Fiorentino, Italy}
\affiliation{INFN, Sezione di Firenze, 50019 Sesto Fiorentino, Italy}

\author{G. Del Pace~\orcidlink{0000-0002-0882-2143}}

\affiliation{Istituto Nazionale di Ricerca Metrologica (INRiM), Sede di Sesto Fiorentino- c/o LENS, Via Nello Carrara, 1 - 50019 Sesto Fiorentino, Italy}
\affiliation{CNR-INO Sede Secondaria di Sesto Fiorentino - c/o LENS, Via Nello Carrara, 1 - 50019 Sesto Fiorentino, Italy}
\affiliation{European Laboratory for Nonlinear Spectroscopy (LENS), Via N. Carrara 1, 50019 Sesto Fiorentino, Italy}
\affiliation{INFN, Sezione di Firenze, 50019 Sesto Fiorentino, Italy}

\author{G. Roati~\orcidlink{0000-0001-8749-5621}}
\affiliation{CNR-INO Sede Secondaria di Sesto Fiorentino - c/o LENS, Via Nello Carrara, 1 - 50019 Sesto Fiorentino, Italy}
\affiliation{European Laboratory for Nonlinear Spectroscopy (LENS), Via N. Carrara 1, 50019 Sesto Fiorentino, Italy}
\affiliation{INFN, Sezione di Firenze, 50019 Sesto Fiorentino, Italy}

\author{F. Marino~\orcidlink{0000-0002-1532-9584}}
	\affiliation{CNR-INO Sede Secondaria di Sesto Fiorentino - c/o LENS, Via Nello Carrara, 1 - 50019 Sesto Fiorentino, Italy}
	\affiliation{INFN, Sezione di Firenze, 50019 Sesto Fiorentino, Italy}  

\date{\today}

\begin{abstract}
Cooperative effects in oscillator networks are often associated with enhanced stability of phase-locked solutions, which increases with system size. We show that the stabilization of persistent currents in annular atomic superfluids with periodic barriers is a concrete manifestation of this phenomenon. Under the simplifying assumption of continuity of atomic flow across identical barriers, the system reduces to a ring of locally coupled Kuramoto-like oscillators. We analytically derive the stability diagram of phase-locked configurations and quantify their robustness to noise and small random initial imperfections, finding excellent agreement with experimental observations. These results are inherent to the ring topology and independent of the specific physical platform.

\end{abstract}

\maketitle

\emph{Introduction.--} Complex networks of simple elements have been extensively studied over the last decades to understand the emergence of collective phenomena in high-dimensional systems. Such macroscopic behaviors result from the interplay between the local dynamics of individual units and the network topology. 

A paradigmatic example is provided by the Kuramoto model of coupled phase oscillators \cite{kura1,kura2,kura3}, in which synchronization phenomena have been extensively investigated for a wide range of coupling topologies, from local (nearest-neighbor) to global \cite{strogatz2000}. Its relative simplicity has enabled tremendous theoretical progress, particularly thanks to the development of analytic techniques to obtain exact mean-field descriptions of the dynamics, such as the Ott–Antonsen ansatz \cite{ott-antonsen}. Beyond its conceptual relevance, the Kuramoto framework provides a common language for synchronization in physical systems. For example, Josephson junction (JJ) arrays, are frequently described by Kuramoto-like models, making them prototypical systems for studying cooperative dynamics in condensed-matter physics \cite{hadley,tsang,mirollo93,watanabe,wiesenfeld95,wiesenfeld98,BaronePaterno1982}.
Despite their theoretical and technological importance, experimental studies of JJ networks with tunable topology remain limited. While individual junctions and small arrays are well studied \cite{Likharev, Tafuri}, assembling large arrays with arbitrary geometries demands advanced nanofabrication, where even minor imperfections can strongly affect system's behavior.

In this context, cold-atom platforms provide a complementary and highly tunable alternative. In contrast to solid-state devices, atomic circuits can be easily assembled and reconfigured in a minimal “Lego-like” fashion, enabling precise tuning of geometry, interactions, and disorder \cite{amico}. A particularly relevant realization is provided by ring-shaped superfluids, the cold-atom analog of superconducting rings. Like their solid-state counterparts, such annular systems support persistent currents in both bosonic and fermionic regimes (see, e.g. \cite{ryu, moulder, dubessy, cai, DelPace22}). 
Recent experiments have exploited this tunability to realize highly controllable atomtronic ring arrays of Josephson junctions using various localized tunneling barriers. 
In Ref.~\cite{aidelsburger}, long-lived supercurrents were observed to form spontaneously from initially incoherent states, with the resulting winding numbers $w$ increasing with the number of barriers $N$. Subsequent works have shown that increasing $N$ also enhances the stability of persistent currents and the system’s ability to support higher-circulation states. The critical circulation beyond which currents decay was found to scale approximately linearly with $N$ \cite{pezze}. Similar behavior occurs in rings with periodic point-like impurities \cite{xhani}, suggesting that these phenomena are largely independent of microscopic coupling details.

Here, we uncover the unifying dynamical mechanism underlying these recent observations. We describe these systems as closed chains of locally coupled Kuramoto-like oscillators, in which phase-locked solutions represent persistent current states. We show that linear stability analysis accurately predicts the critical circulation separating stable and unstable equilibria. Unstable configurations relax into clusters of phase-locked oscillators, corresponding to persistent currents of lower circulation, with the cluster structure reflecting the number of singly-charged vortices generated during decay. 
We analytically derive the stability diagram and the survival probability of persistent currents under noise or imperfect state preparation, demonstrating the cooperative role of ring topology in their formation and stabilization. These results are expected to hold for any physical implementation of the network, including atomic and superconducting platforms. 

\emph{General framework.--} In superconducting and superfluid rings, persistent currents are quantized circulation states, where the phase winds by an integer multiple $w$ of $2\pi$ around the ring \cite{Bloch73}. In ring networks of $N$ nodes ($N \ge 3$), these correspond to the so-called splay (or twisted) states \cite{hadley-splay,wiley,tsang,nichols}, oscillatory solutions in which the phase advances by a constant increment $\phi_w=2\pi w/N$  between neighboring sites, with $1-N\le w \le N-1$, resulting in an overall winding number $w$ across the ring. 

In ring networks of identical nodes with homogeneous local coupling, the existence of such states follows directly from symmetry considerations \cite{stewart1,stewart2}. The network is indeed described by a directed graph [Fig.~1(a)], in which each node is ruled by a dynamical law, and each directed edge indicates how nodes influence one another. The graph is equivariant under all rotations and reflections.
Rotations relabel the nodes $n=1 \, ... \, N$ via $n \mapsto n+1$, and reflections flip the ring across some axis, mapping $n \mapsto -n \,\mathrm{mod}\, N$, consistent with the ring topology. These transformations generate the dihedral symmetry group $\mathbb{D}_N$ of a regular $N$-gon. Since they permute the node labels without altering the governing equations, any equilibrium or periodic solution must be invariant under some subgroup of $\mathbb{D}_N$. In particular, rotation invariance imposes $\theta_n=n \phi_w + c$, $c$ real, corresponding to uniform phase increments between successive nodes. These configurations are indeed associated with the discrete Fourier modes $e^{i\theta_n}$ on a ring array, which form the irreducible representations of the cyclic subgroup $\mathbb{Z}_N \subset \mathbb{D}_N$ \cite{walters}. 

The stability of splay states depends on the details of the local (node) dynamics. Nevertheless, the network rotational symmetry still plays a role in shaping the structure of the stability diagram.
Consider the generic model
\begin{equation} 
f(\dot{\phi}_{n},\dot{\phi}_{n-1}, \phi_n, \phi_{n-1}) = 0\, , 
\end{equation}
where $f$ is a smooth function implementing nearest neighbor coupling on a ring through the phase differences $\phi_n=\theta_{n+1} - \theta_n$ and their derivatives. If the coupling is homogeneous, the system admits splay states $\phi_n = \phi_w$. Introducing a small phase perturbation $\theta_n= n \phi_w  +\eta_n$, with $\eta_n \ll 1$ and linearizing yields
\begin{equation}
a_1 \dot{\delta}_{n} + a_2 \dot{\delta}_{n-1} + a_3 \delta_n + a_4 \delta_{n-1} = 0
\label{lingen}
\end{equation}
where $\delta_n = \eta_{n+1}-\eta_n$ and $a_j(\phi_w)= \partial_{j}f$ are the Jacobian components of $f$ evaluated at $(0,0,\phi_w,\phi_w)$.
Because of the rotational symmetry of the network, the Jacobian is circulant and Eq. \eqref{lingen} is diagonalized by Fourier modes $\delta_n \sim e^{\lambda_k t + ikn}$, with $k = \frac{2\pi m}{N}$, \, $m = 0 ... N-1$. Substituting in Eq. \eqref{lingen}, we obtain the eigenvalues spectrum
\begin{equation}
\lambda_k = - \frac{a_3 + a_4 e^{-i k}}{a_1 + a_2 e^{-i k}}  \, .
\end{equation}
Linear stability requires $\mathrm{Re} \, \lambda_k < 0$ for all modes $k \neq 0$, which is ensured if $g(\phi_w) \equiv (a_1-a_2) (a_3-a_4) >0$.

Since $\phi_w \to 0$ for $N \gg 1$, splay states fill the unit circle more densely, and the stability condition reduces to $g(0)>0$. If the coupling is dissipative, meaning that small phase differences (and their derivatives) are linearly damped favoring synchronization, then $g(0) > 0$, implying that $g$ remains positive in a finite neighborhood around the origin. 
As a consequence, an increasing number of splay states become stable as $N$ increases, and unstable states for lower $N$ can stabilize as they enter the region where $g>0$. This behavior directly illustrates the cooperative stabilization effect.

\begin{figure}
\begin{center}
\includegraphics*[width=1.0\columnwidth]{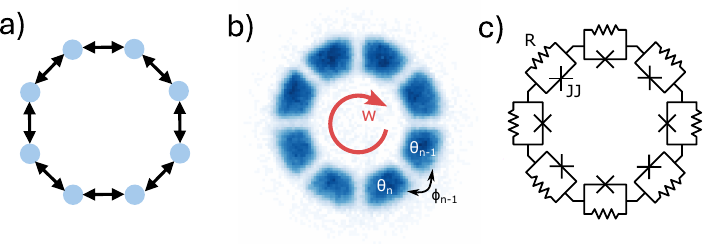}
\end{center} 
\caption{(a) A directed graph representation of a bidirectional ring network of $8$ identical nodes. (b) In-situ density distribution of our annular atomic superfluids with $8$ periodic barriers. c) Equivalent RSJ circuit representation, in which each node is modeled as an ideal Josephson junction shunted by a resistor (see \cite{SM}).}
\label{net}
\end{figure}

\emph{Superfluid ring network.--} The physical system under consideration consists of a ring-shaped atomic superfluid interrupted by an array of $N$ barriers (JJ-links) [see Fig. \ref{net}(b)]. An equivalent circuit representation is shown in Fig.~\ref{net}(c); unlike a standard series connection, each pair of adjacent junctions shares a common superfluid (superconducting) region, resulting in a coupling mechanism distinct from those previously studied in the literature \cite{hadley,mirollo93,wiesenfeld95,wiesenfeld98}. A minimal network model can be constructed by generalizing the standard RSJ approach \cite{WCstewart,McCumber} (see Suppl. Mat. \cite{SM}). The resulting equations read
\begin{equation}
\dot{\theta}_{n+1}- 2 \dot{\theta}_n + \dot{\theta}_{n-1} = \sin(\theta_n - \theta_{n-1}) + \sin(\theta_n - \theta_{n+1}) \, 
\label{eq1}
\end{equation} 
Apart from the diffusive coupling of the time derivatives, Eq. (\ref{eq1}) corresponds to the Kuramoto–Sakaguchi model for identical locally-coupled oscillators in a ring topology \cite{bolo,roy,denes}. Similar to that model, Eq. (\ref{eq1}) is also a gradient system, meaning that limit cycles or more complex solutions are not allowed. The possible attractors thus restrict to fixed points $\{\dot{\theta}_n \}=0$, which correspond to phase-locked configurations defined by the recurrence relations
\begin{eqnarray}
\theta_{n+1} - \theta_n = \theta_n - \theta_{n-1} + 2 m_n \pi \; \label{eq3a} \, , \,\,\,\, n = 1 \, . \,.\, . \,\, N \\ 
\theta_{n+1} - \theta_{n-1} = (2 m_n +1) \pi  \label{eq3b}  \, , \,\,\,\,  n = 1 \, . \,.\, . \,\, N \ ,
\end{eqnarray}
where $m_n$ are site-dependent integers.
\begin{figure}
\begin{center}
\includegraphics*[width=1.0\columnwidth]{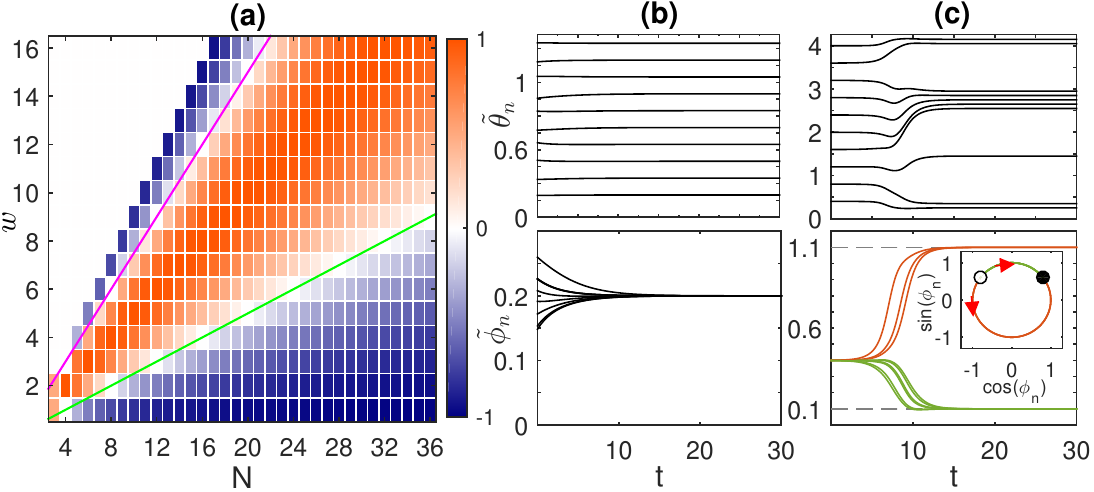}
\end{center} 
\caption{(a) Eigenvalues $\lambda$ of persistent current states $C_{w}^0$ ($0 < w < N$), as a function of $N$. Blue (red) regions where $\lambda$ is negative (positive) indicate stable (unstable) states. Solid lines represent the stability boundaries ($\lambda=0$) given by Eq. (\ref{eq7}): $w=N/4$ (green) and $w=3 N/4$ (magenta). (b-c) Time traces of $\tilde{\theta}_n=\theta_n/2 \pi$ for $N=10$ (upper panel) and $\tilde{\phi}_n=\phi_n/2 \pi$ (lower panel), (b) near the stable state $C_{2}^0$ and (c) near the unstable state $C_{4}^0$. Inset: evolution of $\phi_n$ on the unit circle. The system is initialized with random initial conditions around $C_{w}^0$.} 
\label{figu1}
\end{figure}
The circular geometry imposes the constraint $\theta_{n+N} = \theta_n + 2 \pi w$, where $w$ is the winding number that counts the number of phase-slips in the circular array. This yields the condition
\begin{equation}
\sum_{n=1}^{N} \phi_n = 2 \pi w \, .
\label{wind}
\end{equation}
Only configurations $\{\phi_n\}$ satisfying the above condition correspond to physical solutions of the system. Applied to the solutions (\ref{eq3a}), this implies $\phi_n = 2 \pi q/N + 2 \pi l_n$, where the integer $q$ and the site-dependent integers $l_n$ are such that $q+\sum l_n = w$ \cite{note}. 

We identify persistent current states as configurations that result in a current flow across the array, i.e. characterized by $\sin \phi_n \neq 0$ for all $n$ \cite{SM}. Configurations with $q=w$ and $l_n=0$ correspond to splay states $\phi_w$. In superconducting rings of identical JJs, these states minimize the energy in the classical limit \cite{pop,rastelli}.

When $l_n \neq 0$ for some $n$, the phase pattern consists of clusters of phase-locked oscillators
\begin{equation}
\phi_{n+1} = \phi_n = 2 \pi q/N \,\, , \,\, n = 1 \, . \,.\, . \,\,  n_{cl} < N \ , 
\label{cp}
\end{equation}
where $n_{cl}$ is the number of elements in a cluster. These configurations have the structure of a splay state $\phi_q$ occasionally broken by $2\pi$ phase jumps. Physically, they describe persistent current states of circulation $q$, while the integer $v = \sum l_n$ counts the number of $2 \pi$ defects separating successive clusters. As shown below, solutions \eqref{cp} result from a temporary loss of phase-locking when the system starts in an unstable persistent-current state: $v$ vortices nucleate at specific junctions and the remaining phase differences adjust to conserve total topological charge. We introduce the compact notation $C_{q}^v$ to denote these configurations, with $C_{w}^0$ referring specifically to a persistent current (splay) state $\phi_w$ with circulation $w$ and no vortices involved.

\emph{Stability and dynamics.--} We now assess the stability of splay states $C_{w}^0$ in the parameter space ($w$-$N$). This analysis also applies to any configuration $C_{q}^v$, since the corresponding $\phi_n$ differ at most by integer multiple of $2 \pi$ for some $n$, which do not affect the linear stability. Substituting in Eq. \eqref{eq1} the usual perturbed solution $\theta_n = n \phi_w + \eta_n$, with $\eta_n \ll 1$ we obtain an equation of the form \eqref{lingen} with $a_1=-a_2=1$ and $a_3=-a_4=\cos(\phi_w)$. The eigenvalues $\lambda= - \cos \phi_w$ are real and independent of the Fourier mode $k$. Hence, the stability condition reads
\begin{equation}
\cos \phi_w > 0  \Rightarrow \begin{cases}
    0 < \frac{4 w}{N} < 1 \\
    3 < \frac{4 w}{N} < 4 -\frac{4}{N} \, 
    \end{cases}
\label{eq7}
\end{equation}
for $0 \le w \le (N-1)$. Conditions similar to Eq. (\ref{eq7}) were also derived for circular arrays of Bose-Einstein condensates (BECs) \cite{para,Posa}, and for rings of identical locally-coupled Kuramoto oscillators \cite{wiley,denes,roy,mihara1}. Solutions with $w>N/2$ correspond to negative currents, since $\sin \phi_w <0$, and are  equivalent to states $-w$, since $-w \equiv N-w \, (\mathrm{mod}\, N)$. Thus, we focus on the stability boundary $0 < \frac{4 w}{N} < 1$.

The degree of stability/instability of a configuration $C_{w}^0$ is characterized by the eigenvalues $\lambda$, which quantify the rate at which small perturbations decay or grow. The resulting stability diagram in the ($w$,$N$) plane shows that for any circulation $w$, there exists a minimum network size $N_{c}$, above which the configuration becomes stable. This critical size is determined by the stability boundary $N_{c}=4w$ where $\lambda=0$. Equivalently, for fixed $N$ this defines a critical circulation $w_c=N/4$, beyond which persistent currents decay. 
Figure \ref{figu1}(a) also shows that unstable configurations become less unstable as $N$ increases, while for $N>N_c$ the stability of a given configuration increases, approaching the asymptotic value $\lambda \sim -1$ in the limit $N \gg N_c$.  Therefore, in this limit, finite-circulation states $C_{w}^0$ exhibit the same degree of stability as the zero-circulation state, for arbitrary $w$.

An example of the transient dynamics starting from a perturbed initial condition near $C_{w}^0$ is shown in Fig. \ref{figu1}(b-c).
If the state is stable, we observe a rapid return to the equilibrium (see Fig. \ref{figu1}(b)).
In contrast, initializing the system near an unstable state results in a transition to a cluster solution $C_{q}^v$, where $q+v=w$ due to conservation of topological charge. In Fig. \ref{figu1}(c) we show the transition $C_{4}^0 \rightarrow C_{1}^3$ for $N=10$. The initial nearly-regular arrangement of phases breaks into four clusters of phase-locked oscillators 
The dynamics becomes particularly clear when plotting the phase-differences $\phi_n$ of all nodes. Starting into the neighborhood of $\phi_n= 2 \pi w/N=4 \pi/5$ ($w=4,N=10$) each trajectory follows a heteroclinic orbit connecting the unstable fixed point to the stable equilibrium at $2 \pi/10$ (mod $2 \pi$), which for $N=10$ corresponds to a circulation $q=1$. This state is indeed stable, as indicated by $\lambda_n =-\cos(2 \pi/10) <0$. 
\begin{figure}
\begin{center}
\includegraphics*[width=1.0\columnwidth]{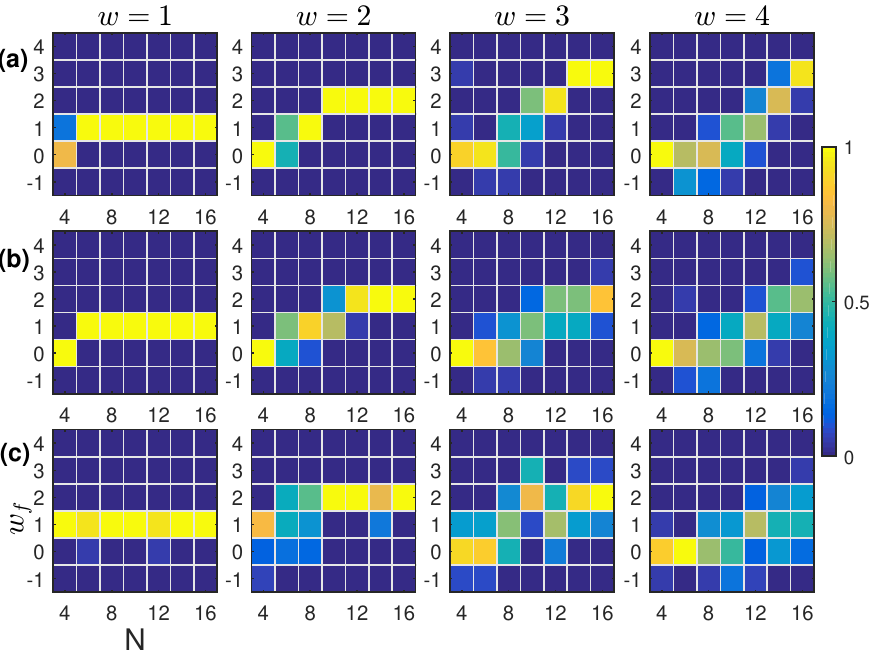}
\end{center} 
\caption{Final-state probabilities $P(w_f \vert w)$ obtained by integrating Eq.~(\ref{eq1}) for $20$ random initial conditions in the vicinity of $C_{w}^0$,
for $w=1,2,3,4$, shown as a function of $N$: (a) deterministic evolution; (b) evolution in the presence of zero-mean spatiotemporal additive Gaussian noise $\zeta(n,t)$ with correlation $\langle \zeta(n',t') \zeta(n,t)\rangle = 2 D \,\delta(n'-n)\delta(t'-t)$ and $D=8.45 \times 10^{-3}$. The winding number is calculated via Eq. \eqref{wind}. (c) Experimental final-state probabilities from $15$-$20$ realizations for each $N$ and $w$. 
}
\label{figu2}
\end{figure}
In the circular phase-space representation (inset), trajectories repelled from the unstable point (open circle) may flow clockwise directly to the nearby stable equilibrium (filled circle) or counterclockwise, reaching the same attractor after a longer excursion. The number of counterclockwise loops corresponds to the number of nucleated vortices and characterizes the final cluster configuration. 

\emph{Theoretical vs experimental results.--} For a given initial $w$ and $N$, the final state depends on the initial conditions and can involve any of the nearby stable attractors. All these possible dynamical outcomes can be characterized through the final-state probability distribution $P(w_f \vert w)$, which specifies the probability that an initial condition near the $C_{w}^{0}$ evolves into a stable (cluster) state with circulation $q=w_f$. The results are shown in Fig. \ref{figu2}(a). When the initial configuration is linearly stable ($N \geq 4w$) the system remains in that state with probability close to 1, so that $w_f=w$. If the initial configuration is unstable, the system relaxes into any of the nearby coexisting stable states, depending on the initial condition. The inclusion of spatiotemporal noise, unavoidable in real systems, induces escape events even from stable states. As a result, the regions where the probability of remaining in the initial state equals $1$ shrinks, and the overall distribution broadens, as shown in Fig. \ref{figu2}(b).

We now compare these results with those obtained experimentally in a superfluid ring with periodic barriers. In the experiment, we produce a Bose–Einstein condensate of $^{6}$Li molecules confined in a repulsive, ring-shaped hard-wall potential generated by a digital micromirror device~\cite{DelPace22}. 
Persistent currents with different winding numbers $w$ are excited using a phase-imprinting technique~\cite{DelPace22}. The experimental $w$ is extracted via an interferometric probe (see \cite{SM}). 
To realize the atomic JJ array, after the phase-imprinting, we superimpose up to $N = 16$ equally spaced optical barriers onto the ring superfluid. Barrier widths and heights are tailored to ensure operation in the tunneling regime \cite{SM}.

The stability of the persistent current is extracted by evaluating the final $w_f$ after a given observation time \cite{SM}. In this way, we determine the final-state probability $P(w_f \mid w)$ as a function of $N$ and we compare it with the calculated values. The results are shown Fig.~\ref{figu2}(c). A better agreement is found with the numerical results in Fig.~\ref{figu2}(b), where a small amount of spatiotemporal noise has been added to the system. This is not unexpected as technical noise and fluctuations of thermal origin are intrinsic to experiments and affect the dynamics.
The robustness of a persistent current state $C_{w}^{0}$ against noise or random initial imperfections can be quantified by the survival probability $P_{\rm surv}(w,T)$, defined as the probability that the system trajectories remain within an N-dimensional ball of radius $r$ around the state after an observation time $T$. Assuming small isotropic Gaussian perturbations with initial variance $\sigma^2(0)$, and linear evolution around the state $C_{w}^{0}$, the survival probability takes the exact closed form (see Suppl. Mat. \cite{SM})
\begin{equation}
P_{\rm surv}(T)
=\frac{\gamma\left(\tfrac{N}{2},\tfrac{r^2 e^{-2 \lambda T}}{2\sigma^2(0)}\right)}{\Gamma(\tfrac{N}{2})},
\label{probsurv}
\end{equation}
where $\gamma(s,x)$ is the lower incomplete gamma function and $\Gamma(s)$ the gamma function. Eq. \eqref{probsurv} provides a rough estimation of the probability that the perturbations remain small enough not to push the system out of the basin of attraction of the initial state \cite{note-ba}. For unstable fixed points $\lambda>0$, the survival probability always decays to zero as $T \rightarrow \infty$, whereas for stable states $\lambda<0$, it converges to a finite limit, which approaches unity if $r$ is large compared to the stationary noise amplitude. For any $w$ and $T>0$, $P_{\rm surv}(T) \rightarrow 1$ for $N \gg 1$, and the approach is exponentially fast $1-P_{\rm surv}(T) \sim \exp(-N)$. This implies an enhanced robustness of persistent current states against fluctuations as $N$ increases, due to the cooperative effect. 

A direct experimental measurement of $P_{\rm surv}$ would require resolving the dynamics shortly after the initial evolution, where the linear approximation is still valid. On the other hand, the final-state probability distributions provide the experimentally accessible indicator $\mathcal{S}=\langle w_f \rangle/w$, where $\langle w_f \rangle$ is the average final circulation calculated over multiple realizations. This quantity, previously used in Ref. \cite{pezze} to characterize the robustness of persistent-current states, carries some information about $P_{\rm surv}$. 
A simple way to illustrate this connection is to note that if a state with winding number $w$ survives with probability $P_{\rm surv}$, it has a probability $1 - P_{\rm surv}$ to evolve into a different circulation state. Hence, we can write 
\begin{equation}
\langle w_f \rangle \approx w P_{\rm surv} + \langle w_f \rangle_{esc} (1 - P_{\rm surv}) \, ,
\label{toy}
\end{equation}
where $\langle w_f \rangle_{esc}$ is the conditional average of the final winding number, computed only over the realizations that leave the basin of attraction of the initial state. 
Note that, for stable states, the main contribution to $\langle w_f \rangle$ comes from $P_{\rm surv}$, while for unstable states $\langle w_f \rangle \approx \langle w_f \rangle_{esc}$. The two contributions become comparable close to the stability boundary, where states are marginally stable.

\begin{figure}
\begin{center}
\includegraphics*[width=1.0\columnwidth]{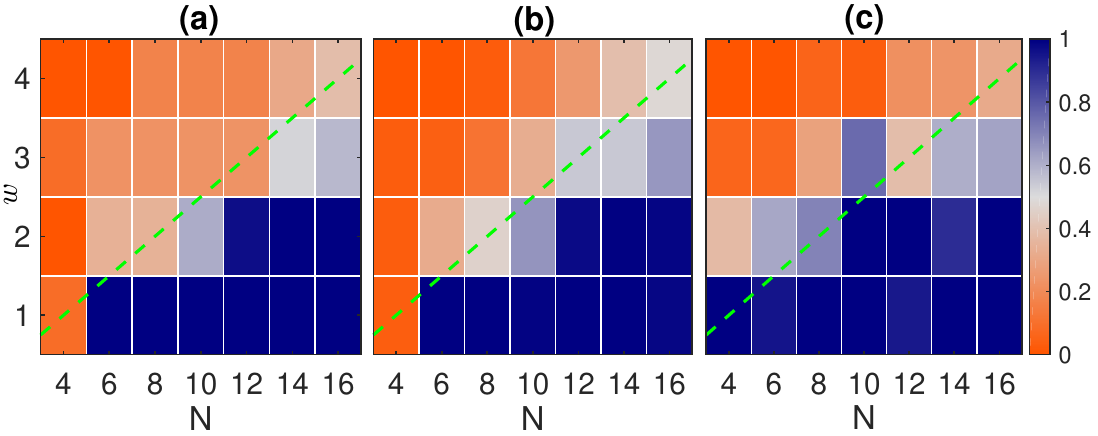}
\end{center} 
\caption{Robustness of persistent current states $C_{w}^0$ ($0 < w < N$), as a function of $N$ quantified by the indicator $\mathcal{S}$ (colormap). Blue (red) regions indicate more (less) robust states. Dashed lines marks the stability boundary $w=N/4$. (a) $\mathcal{S}$ calculated from Eq. \eqref{toy}, using $\langle w_f \rangle_{esc} = \frac{N}{2\pi} \arccos(-\Lambda)$, $\sigma=6\times10^{-2}$, $r=\sigma$ and $T=3$; (b) $\mathcal{S}$ computed from numerical solutions of Eq. \eqref{eq1}, with parameters as in Fig. \ref{figu2}(b), averaged over $20$ realizations; (c) $\mathcal{S}$ extracted averaging over $15$-$20$ experimental realizations for each $N$ and $w$.}
\label{figu3}
\end{figure}

An exact analytical expression for $\langle w_f \rangle_{esc}$ is difficult to obtain, as it depends on the nonlinear evolution of the system.
Following a mean-field projection of the unstable manifold dynamics, we assume $\langle w_f \rangle_{esc} = \frac{N}{2\pi} \arccos(-\Lambda)$, where $\Lambda$ is the average of curvature of accessible stable states (see Suppl. Mat. \cite{SM}). In Fig. \ref{figu3}(a) we plot $\mathcal{S}$ obtained from Eq.~\eqref{toy}. We observe that all configurations above the critical threshold $N_c > 4 w$ are mostly robust, according to the stability condition \eqref{eq7}. In Fig.~\ref{figu3}(b) we show $\mathcal{S}$ calculated numerically from Eq.~\eqref{eq1}.  The agreement is good, even near the stability boundary where the differences between the analytic and numerical $\langle w_f \rangle_{esc}$ are larger. The experimental $\mathcal{S}$ 
are plotted in Fig. \ref{figu3}(c) and well agree with the theoretical predictions. The most significant difference occurs for $(w,N)=(1,4)$, which is predicted to be marginally stable, but appears stable in the experiment. 
This is related to the reduced accuracy of representing the continuous annular superfluid as a discrete network at very low $N$. These results demonstrate the cooperative effect of the ring topology.

A counterintuitive consequence of this effect is that the probability of spontaneously generating persistent currents from random initial conditions increases with system size. In a recent experiment \cite{aidelsburger}, $N$ initially uncorrelated Bose–Einstein condensates were merged into a ring-shaped optical trap. The resulting winding-number distribution was found to broaden as $\sqrt{N}$, leading to an enhanced probability of forming states with larger $w$, consistently with the geodesic rule. For phase differences uniformly distributed over $[-\pi,\pi)$ the expected width of the distribution is $\sigma_w =\sqrt{N}/(2 \sqrt{3}) \sim 0.29 \sqrt{N}$ \cite{aidelsburger,janson}. Our model \eqref{eq1} reproduces the $\sqrt{N}$-scaling with high accuracy. We fitted the numerical $\sigma_w$ to $A(N)=\sigma_0 N^{b}$, and obtained $b=0.505 \pm 0.009$ and $\sigma_0=0.180\pm0.002$ (see \cite{SM}). The smaller prefactor $\sigma_0$ suggests weak correlations of the final phases over domains of a few sites, rather than a uniform distribution over the full circle.
This is in excellent agreement with the predicted $\sim 0.2\sqrt{N}$ scaling of basin volumes in locally coupled Kuramoto rings \cite{wiley}, and can be related to cooperative stabilization via established basin–eigenvalue relations \cite{ochab,mihara2}. As the system size increases, the eigenvalues of stable splay states become more negative and their basins of attraction expand, which enhances both the spontaneous formation of persistent currents and their robustness against noise and imperfections. These results establish a direct link between the geodesic rule and the underlying network dynamics (see \cite{SM}).

\emph{Conclusions.--} We have shown that the stabilization of persistent currents in annular atomic superfluids with periodic barriers is the result of a cooperative phenomenon in Kuramoto-like ring networks. 
%
As the number of junctions $N$ grows, persistent currents become more robust to noise and imperfections and are more likely to form spontaneously. This behavior is an intrinsic property of ring networks and is therefore largely independent of the specific physical platform.
Our measurements, together with experiments in \cite{aidelsburger,pezze,xhani}, demonstrate that cooperative effects are robust in the presence of the small, but experimentally unavoidable, disorder that breaks the rotational symmetry of the barrier configuration. We expect these effects to persist even under moderate disorder, providing an effective error-correction mechanism for the generation and stabilization of persistent currents in non-ideal conditions \cite{new}. 
These insights not only account for current observations in cold-atom systems, but also pave the way for exploiting cooperative effects in broader contexts, including the realization of robust superconducting circuits.


\emph{Acknowledgements.--} We thank Jean Dalibard, Giulio Nesti and Luca Pezz\'e for stimulating discussions and useful comments on the manuscript. 
G.R. acknowledge financial support from the PNRR MUR project PE0000023-NQSTI and from the Italian Ministry of University and Research under the PRIN2017 project CEnTraL and project CNR-FOE-LENS-2024. 
The authors acknowledge support from the European Union - NextGenerationEU within the “Integrated Infrastructure Initiative in Photonics and Quantum Sciences" (I-PHOQS). 
The authors acknowledge funding from INFN through the RELAQS project. 
This publication has received funding under the Horizon Europe programme HORIZON-CL4-2022-QUANTUM-02-SGA (project PASQuanS2.1, GA no.~101113690) and Horizon 2020 research and innovation programme (GA no.~871124).
\section{Appendix}
\appendix

\section{Superfluid ring network model}

The physical system under consideration consists of an array of $N$ barriers in a ring-shaped atomic superfluid. This setup resembles a circular array of Josephson junctions connected in series, but with important differences: each pair of adjacent junctions shares a common superfluid region, and the array does not couple to a common load (see e.g. \cite{hadley}). This distinction leads to a coupling mechanism that differs from the global configuration, typically studied in the literature \cite{hadley,mirollo93,wiesenfeld95,wiesenfeld98}. 
A minimal model can be constructed by generalizing the standard RSJ approach \cite{WCstewart,McCumber}. In analogy with superconducting junctions, the total current across each barrier is written as the sum of a superfluid component and a normal (dissipative) one, $I_n = I_c \sin(\theta_n - \theta_{n-1}) + \hbar(\dot{\theta}_n - \dot{\theta}_{n-1})/R$, where here $\theta_n$ is the phase in the $n$-th superfluid region, $I_c$ is the critical current and the phenomenological mass-resistance $R$ encodes dissipative processes at the barrier. In the atomic superfluid Josephson junctions considered, the dissipative current associated with the resistance in the circuit analog originates from excitations such as phonons and vortices \cite{delpace_prl,burchianti2018,xhani2020}. 
We assume identical junctions (i.e a single $I_c$ and $R$) and neglect additional losses and capacitive effects. Imposing current continuity between adjacent superfluid regions, $I_{n-1}=I_{n}$ for all $n = 1 \, . . . \, N$, yields 
\begin{equation}
\dot{\theta}_{n+1}- 2 \dot{\theta}_n + \dot{\theta}_{n-1} = \sin(\theta_n - \theta_{n-1}) + \sin(\theta_n - \theta_{n+1}) \, 
\label{eq1}
\end{equation}
where all currents have been normalized to $I_c$, and we introduced the dimensionless time $t$ related to the physical time $t_{ph}$ via $t = (R I_c / \hbar) t_{ph}$.

\section{Survival probability and robustness indicator}

The robustness of a persistent current state against noise or small random initial imperfections can be quantified through the survival probability $P_{\rm surv}(w,T)$, defined as the probability that the system remains in a state with circulation $w$ after an observation time $T$. This probability can be estimated analytically under the following simplifying assumptions. We consider the linearized dynamics of the perturbation vector $\{ \delta_n\}$ around the state $C_{w}^{0}$. For isotropic Gaussian initial perturbations, the components $\delta_n(0)$ are independent, zero-mean normal variables with variance $\sigma^2(0)$. In the linear regime, after time $T$, $\delta_n(T)$ remain independent zero-mean Gaussians with variance $\sigma^2(T)=\sigma^2(0) e^{2 \lambda T}$. We define the survival of a state with winding number $w$ up to time $T$ by requiring that the perturbations remain within an N-dimensional ball of radius $r$, i.e. $\sum_{n=1}^N \delta_n(T)^2 < r^2$.
Introducing the normalized variables $Z_n=\delta_n(T)/\sqrt{\sigma(T)}$, the survival condition becomes $\sum_{n=1}^N Z_n^2 < r^2/\sigma^2(T)$.
Since this sum is $\chi^2$-distributed with $N$ degrees of freedom, the survival probability takes the exact closed form
\begin{equation}
P_{\rm surv}(T) \equiv P\left(\chi^2_N < r^2/\sigma^2(T)\right)
=\frac{\gamma\left(\tfrac{N}{2},\tfrac{r^2 e^{-2 \lambda T}}{2\sigma^2(0)}\right)}{\Gamma(\tfrac{N}{2})},
\label{probsurv}
\end{equation}
where $\gamma(s,x)$ is the lower incomplete gamma function and $\Gamma(s)$ the gamma function.

As discussed in the main text, $P_{\rm surv}$ is related to the robustness indicator $\mathcal{S}=\langle w_f \rangle/w$ through the relation \begin{equation}
\langle w_f \rangle \approx w P_{\rm surv} + \langle w_f \rangle_{esc} (1 - P_{\rm surv}) \, ,
\label{toy}
\end{equation}
where $\langle w_f \rangle_{esc}$ is the conditional average of the final winding number, computed only over the realizations that leave the basin of attraction of the initial state. 
Note that, for stable states, the main contribution to $\langle w_f \rangle$ comes from $P_{\rm surv}$, while for unstable states $\langle w_f \rangle \approx \langle w_f \rangle_{esc}$. The two contributions become comparable close to the stability boundary, where states are marginally stable.
Eq. \eqref{toy} is semi-phenomenological, since $\langle w_f \rangle_{esc}$ must be determined through heuristic reasoning (see next section) or numerical estimation. It nevertheless has the advantage of making explicit the link between $\mathcal{S}$ and the eigenvalues through $P_{\rm surv}$. While global dynamical properties generally cannot be inferred from local linear features such as eigenvalues, \eqref{eq1} is a gradient system whose dynamics consist of heteroclinic trajectories connecting fixed points on a compact phase space ($N$-torus). In such systems the influence of eigenvalues extends far beyond the immediate vicinity of the equilibrium \cite{mihara1,mihara2}.

\section{Estimating $\langle w_f \rangle_{esc}$ via a mean-field toy model}

To predict the $\langle w_f \rangle_{esc}$ reached from an unstable state $w$ in a system with many nearby stable states/potential wells, we exploit a mean-field description of the unstable manifold dynamics. The gradient flow projects trajectories onto a manifold that is approximately one–dimensional and parameterized by an effective curvature. What matters is not which particular wells were crossed, but the average stabilizing curvature experienced along the unstable manifold.

To this aim, we consider only the physically stable wells $w < N/4$, whose eigenvalues $\lambda(w_i)$ are negative. Starting from $w$, we include successive nearby stable states until the cumulative stabilizing curvature overcomes the local instability, i.e. until that
\begin{equation}
\lambda(w) + \sum_{i=1}^{k} \lambda(w_i) < 0.
\label{ec}
\end{equation}
This identifies the minimal set of wells that the unstable manifold can effectively probe, before the dynamics is forced to collapse into the final basin. Rather than tracking which specific wells are crossed, we replace this set by its average curvature
\begin{equation}
\Lambda = \frac{1}{k} \sum_{i=1}^{k} \lambda(w_i) \, ,
\end{equation}
which represents the mean stabilizing influence experienced along the unstable direction. The final attractor is then taken as the well whose curvature matches this average,
\begin{equation}
\langle w_f \rangle_{\rm esc} = \frac{N}{2\pi} \arccos(-\Lambda) \, .
\end{equation}

Conceptually, this procedure reflects a mean-field projection of the dynamics along the unstable manifold. The trajectory initially experiences the local positive curvature $\lambda(w)$, but is progressively stabilized by the surrounding wells. Once the cumulative stabilizing curvature exceeds the initial instability, the flow is inevitably captured by a basin. The mean-field approximation assumes that the detailed sequence of wells visited is less important than the net stabilizing effect they collectively exert. Mathematically, this is equivalent to replacing the heterogeneous local curvatures along the sampled unstable manifold by their average $\Lambda$, which then determines the final state.

This approach naturally predicts several features of the dynamics. Larger $\lambda(w)$ or smaller stabilizing curvatures shift the final attractor farther along the manifold (see Fig. $2$ of the main text), while the set of accessible wells is automatically restricted by Eq. \eqref{ec}. Noise or initial fluctuations only affect which wells contribute to the average, without altering this heuristic description. 

\section{Spontaneous emergence of persistent current states}

In this section we briefly review the experimental findings and theoretical interpretation of the results in Ref. \cite{aidelsburger} and discuss how they connect to the collective stabilization mechanism.

Ref. \cite{aidelsburger} provided a detailed study of the out-of-equilibrium dynamics of $N$ uncorrelated condensates after connecting them in a ring-shaped optical trap. They observed the spontaneous formation of long-lived persistent currents and reported that the distribution of the resulting winding numbers broadens with increasing $N$, in quantitative agreement with the so-called geodesic rule. 

According to the geodesic rule, the system selects the winding number $w$ corresponding to the phase configuration that minimizes the total variation of phase differences, and therefore the total kinetic energy. Before merging, each condensate segment carries an independent random phase. Upon connection, the phase must become single-valued around the ring, so that the final winding number $w$ is determined by the the total accumulated phase around the ring, given by the sum of the local phase differences $\phi_n$ between neighbors (see Eq. $(7)$ of the main text). If the initial phases are independent and uniformly distributed, the resulting distribution of $w$, $P(w)$, is given exactly by the Euler–Frobenius distribution \cite{janson}, whose width grows with $\sqrt{N}$. In the large-$N$ limit this distribution approaches a Gaussian with standard deviation $\sigma_{w} \sim \sqrt{N}$. 

In general, if $\phi_n$ are independent variables, randomly distributed with zero-mean and finite variance $\sigma^2_{\phi}$, then 
\begin{equation}
\sigma_{w} = \frac{1}{2 \pi} \sqrt{\sum_{n=1}^N \sigma^2_{\phi} }= \frac{\sigma_{\phi}}{(2 \pi)} \sqrt{N} \, .
\end{equation}
This is universal, independent of the exact distribution. Consequently, as $N$ increases, $P(w)$ becomes wider and the probability of forming states with nonzero $w$ increases.

In dissipative nonlinear systems, the probability of reaching a given final state from random initial conditions is determined by the basin volume of that state with respect to the chosen initial-condition distribution. The observed winding-number statistics therefore reflect the fraction of sampled initial conditions that fall into the basin of attraction of each phase-locked state. In locally coupled Kuramoto rings, Wiley, Strogatz, and Girvan \cite{wiley} showed numerically that the probability of converging to a splay state of winding $w$ (and thus its basin volume) is well described by a Gaussian distribution whose width scales as $\sim 0.2 \, \sqrt{N}$. Although the dynamics and basin structure are fully deterministic, this behavior was explained statistically through the central limit theorem, establishing a first direct connection between the geodesic-rule and the basin volume scaling \cite{wiley}.

Subsequent works demonstrated that in these networks basin sizes of splay states are strongly correlated with their linear stability eigenvalues \cite{ochab}. In particular, Mihara et al. showed that the sum of the eigenvalues of stable $w$-states behaves similarly to their linear basin size (see Ref. \cite{mihara2} and Fig. 3 in Ref. \cite{delabays}) and reproduces exactly the $\sqrt{N}$ scaling obtained by numerical simulations in Ref. \cite{wiley}. This indicates that the eigenvalue spectrum somehow encode information on how the basin volumes of splay states are distributed as a function of $N$.

\begin{figure}
\begin{center}
\includegraphics*[width=1.0\columnwidth]{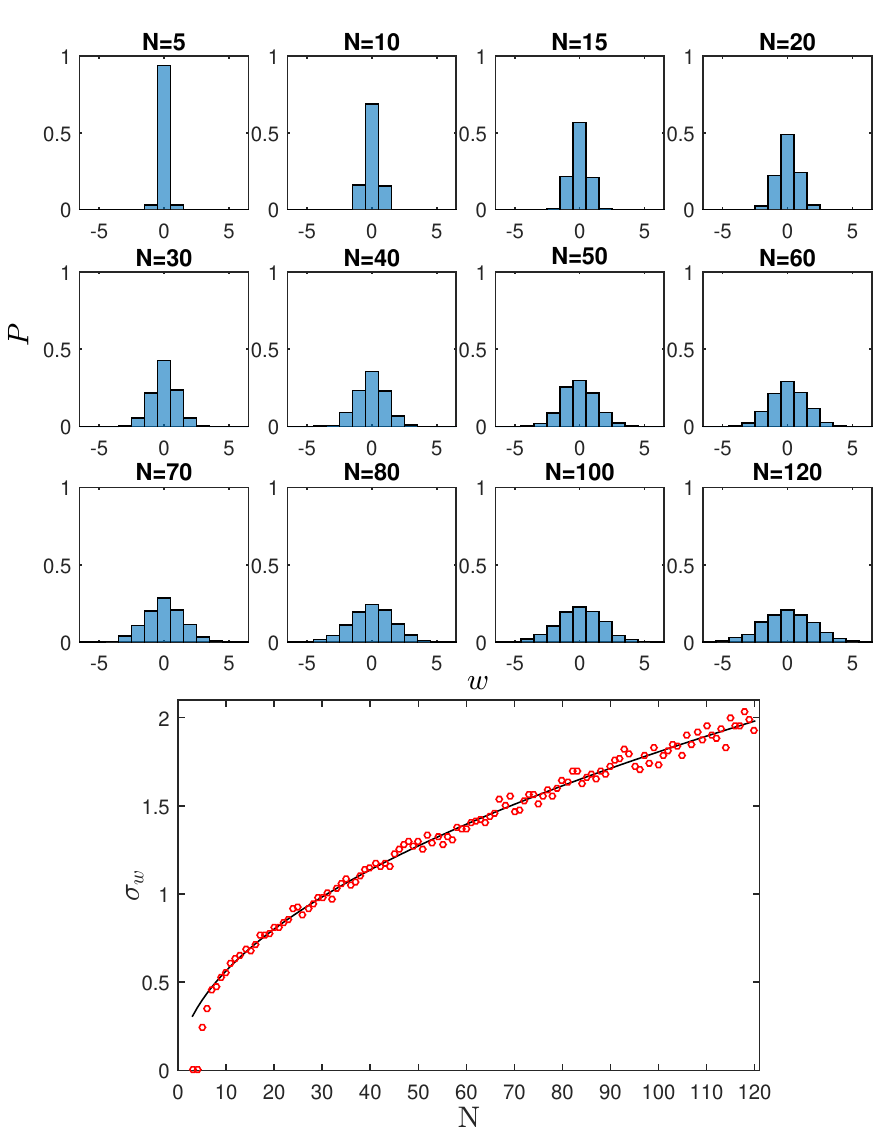}
\end{center} 
\caption{Final-state probabilities $P(w)$ for different $N$ calculated by integrating Eq. (4) of the main text over a numerical time $t_{f}=80$, starting from $1000$ different initial conditions with $\theta_n(0)$ distributed uniformly in the interval $[0,2\pi)$. Lower panel: measured standard deviation of the probability distributions as a function of $N$. The winding number is calculated via Eq. (7). }
\end{figure}

While in general global dynamical properties such as basin volumes cannot be inferred from purely local characteristics like Jacobian eigenvalues, in locally-coupled Kuramoto rings, the dynamics is strongly constrained by two non simply-connected (circular) topologies: one in real space and another arising from the compactness of the phase variables. As discussed in Refs. \cite{wiley,mihara2}, these systems exhibit gradient dynamics on a compact manifold, with (heteroclinic) trajectories flowing monotonically over a potential surface and asymptotically connecting fixed points. In such systems, the influence of the eigenvalues extend over large distances from the equilibrium state.

The cooperative stabilization mechanism discussed in this work, together with the mathematical results of Refs. \cite{wiley,ochab,mihara2}, thus provide an interesting link between the geodesic rule and network dynamics.
As the system size increases, the eigenvalues of stable splay states become more negative, causing their basins of attraction to expand. This cooperative basin expansion explains not only the increased probability of spontaneous formation of such states, but also their enhanced robustness against noise, disorder, and imperfections, independently of the specific preparation protocol. 

We conclude this section by showing that our network model (see Eq. $(4)$ of the main text) qualitatively reproduce the experimental observations of Ref. \cite{aidelsburger}. In Fig. S1 we report the probability distributions for different network sizes obtained from numerical integration of Eq. ($4$) of the main text over $1000$ realizations of initial phase differences randomly distributed in the interval $(0,2\pi]$. 

While perfect quantitative agreement with experimental distributions is not expected, especially at small $N$ due to the limited accuracy of representing the continuous annular superfluid as a discrete network, we nonetheless observe a broadening of the distributions as 
$N$ increases. This leads to an enhanced probability of observing states with higher winding numbers, consistent with experimental observations. The standard deviation of the winding number exhibits the expected $\sqrt{N}$-scaling, also observed in \cite{aidelsburger} and explained in terms of the geodesic rule. For configurations with phase differences uniformly distributed over the full interval $[-\pi,\pi)$ the geodesic rule predicts $\sigma_w =\sqrt{N}/(2 \sqrt{3}) \sim 0.29 \sqrt{N}$, in agreement with the Euler–Frobenius distribution \cite{janson,aidelsburger}.
We fitted the numerical $\sigma_w$ to $A(N)=\sigma_0 N^{b}$ calculated from numerical integration of Eq. (4) of the main text, and obtained $b=0.505 \pm 0.009$ and $\sigma_0=0.180\pm0.0015$. This is in excellent agreement with the predicted $\sim 0.2\sqrt{N}$ scaling of basin volumes in locally coupled Kuramoto rings \cite{wiley}. A similar $\sqrt{N}$-scaling has been obtained in Ref. \cite{denes}. The lower value $\sigma_0$ compared to the Euler–Frobenius prediction indicates that the random phases in our system are not uniformly distributed over the full circle, but are instead constrained, on average, to a smaller interval. This suggests that neighboring phases are partially correlated over domains extending beyond a single node, while remaining random and effectively independent beyond the size of each domain.

\section{Experimental methods}

\begin{figure}[t]
    \centering
    \includegraphics[width=0.7\linewidth]{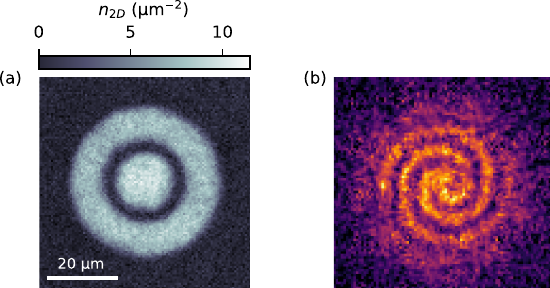}
    \caption{
    \textbf{(a)} Experimental superfluid configuration consisting of a superfluid ring and an inner disk. The image is averaged over 10 experimental realizations. The colormap shows the two-dimensional density integrated along the $\hat{z}$ direction, $n_\mathrm{2D}$. 
    \textbf{(b)} Observed interferogram after time-of-flight expansion for the case of $w=2$ before introducing the barriers.
    }
    \label{fig:experimentalfigure}
\end{figure}

We prepare the superfluid sample by evaporating a balanced mixture of the two lowest hyperfine states of $^6$Li atoms at $\SI{832}{G}$, close to their Feshbach resonance, in an elongated elliptic optical dipole trap. After evaporation, the magnetic field is ramped to $\SI{702}{G}$, bringing the system into the molecular Bose--Einstein condensate (BEC) regime.
The gas is then transferred to the final configuration by adiabatically ramping up a repulsive $\mathrm{TEM_{01}}$-like optical potential at $\SI{532}{nm}$, which provides strong confinement along the $\hat{z}$ direction, with a trapping frequency $\omega_z = 2\pi \times 383(2)\,\mathrm{Hz}$. 
In the $\hat{x}$--$\hat{y}$ plane, a repulsive hard-wall cylindrically symmetric potential is simultaneously turned on to confine the sample within a circular region. This circular box potential is projected onto the atomic cloud using a digital micromirror device (DMD). Both the magnetic-field sweep and the optical-potential ramp are performed over the last $\SI{100}{ms}$ of the evaporation sequence.
A residual radial harmonic potential of $\SI{2.5}{Hz}$ is present in the plane, originating from the combined effect of the anti-confinement induced by the $\mathrm{TEM_{01}}$ beam in the horizontal plane and the confining curvature of the magnetic field used to tune the Feshbach field. Over the box-trap radius, this weak confinement has a negligible effect, resulting in a nearly homogeneous density distribution.
An additional circular repulsive potential is then raised to create an external superfluid ring surrounding an inner disk (see Fig.~\ref{fig:experimentalfigure}a). The superfluid ring is characterized by an inner radius $R_{\rm in} = 11.7 \pm 0.2~\mu\mathrm{m}$ and an outer radius $R_{\rm out} = 20.6 \pm 0.2~\mu\mathrm{m}$. The number of molecules in the ring is $N_p = (6.9 \pm 0.3)\times10^4$, where the uncertainty reflects the shot-to-shot reproducibility.
In this trapping and interaction configuration, the Fermi energy and the chemical potential can be estimated as \cite{DelPace22}:
\begin{align}\label{eq:mubec}
E_F &= 2\hbar\left[\frac{\hbar\omega_z N_p}{m(R_\mathrm{out}^2-R_\mathrm{in}^2)}\right]^{1/2},\\
\mu &= \left( \frac{3}{2} \frac{\hbar^2 \omega_z N_p a_M}{\sqrt{m}\,(R_{out}^2 - R_{in}^2)} \right)^{2/3}, 
\end{align}
where $a_M = 0.6\,a_s$ is the molecule--molecule scattering length, with $a_s$ being the atom--atom scattering length.
The gas is characterized by a $E_F/h = \SI{7.2 \pm 0.2}{kHz}$ and $\mu/h = \SI{850 \pm 20}{Hz}$, with a value of $1/k_Fa_s = \SI{3.8\pm 0.1}{} $, where $k_F = \sqrt{2mE_F}/\hbar$ is the Fermi wave vector.

After preparing a stationary superfluid ring, the circulation state along the ring is controlled using a phase-imprinting technique, following the procedure in \cite{phase_imprinting_pra,DelPace22}. Specifically, we apply an optical potential generated with the DMD, whose intensity increases linearly along the azimuthal direction within the superfluid ring region, $V(\theta) = V_0 \theta$.
By applying this potential for an imprinting time $t_I \ll \hbar/\mu$, its only effect on the cloud is to imprint a phase that increases linearly with the azimuthal direction, $\phi(\theta) = V(\theta)t_I/\hbar$, resulting in a total phase winding $\Delta\phi_I = V_0 t_I/\hbar$, which is proportional to the controllable pulse duration. This protocol enables the preparation of controlled, high-fidelity quantized circulation states in the ring.
Following the imprinting, we wait $300\,$ms to allow the system to relax to equilibrium and for any density excitations induced during the imprinting process to decay. 

We then introduce the desired number $N$ of Gaussian-shaped barriers, realized using the DMD, experimentally reproducing an array of Josephson junctions with phase winding $w$ and number of barriers $N$. 
Each barrier is positioned at equally spaced azimuthal angles and has a height $U_0 = (1.3 \pm 0.2)\,\mu$ and a $1/e^2$ width $\sigma = (1.2 \pm 0.2)\,\xi$, where $\xi$ is the superfluid healing length ($\xi \simeq 0.68~\mu\mathrm{m}$). These parameters ensure operation in the tunneling regime. The values and uncertainties of the barrier parameters correspond to the mean and standard deviation over the set of barriers, respectively.
The barriers are ramped on over $1~\mathrm{ms}$ to minimize the excitation of unwanted collective modes.

The value of $w$ in each experimental realization is obtained through an interferometric measurement \cite{corman2014,DelPace22}. Specifically, we let the outer ring interfere with the inner disk during a time-of-flight expansion, in which the optical potentials are abruptly switched off.
In the case $w \neq 0$, the resulting interference pattern exhibits a spiral structure. This arises from the linear increase of the relative phase between the ring and the disk as a function of the azimuthal angle, which leads to a shift of the interference fringes at each angular position.
The interference pattern allows for a direct determination of $w$, since the number of spiral arms corresponds to the value of the circulation (see Fig.~\ref{fig:experimentalfigure}b). This behavior is expected considering the periodicity of the interference fringes after an increase of the relative phase of $2\pi$. As a result, the pattern repeats itself $w$ times over a full azimuthal rotation, giving rise to $w$ spiral arms. The direction of rotation of the arms is instead linked to the sign of the circulation state. The labeling of positive and negative circulation is chosen arbitrarily.


\begin{thebibliography}{99}
\bibitem{note} Solutions (\ref{eq3b}) satisfying the periodic boundary conditions take the form $\phi_n = m_n \pi$ if $N$ is odd, or $\phi_n =(-1)^n a + m_n \pi$, with $a$ real if $N/2$ is even. No solutions of this kind exists for $N/2$ odd.
\bibitem{note-ba} The basin boundaries are determined by the location of the saddle points separating neighboring phase-locked configurations. In our system these saddles lie at finite distances from stable states, where the linearized dynamics used in deriving $P_{\rm surv}$ is no longer valid. Therefore \eqref{probsurv} is accurate only within a neighborhood strictly inside the basin.
\end{thebibliography}
\end{document}